# Impact of High Penetration of Inverter-based Generation on Electromechanical Wave Propagation in Power Grids

Shutang You

*Abstract*— A power system electromechanical wave propagates from the disturbance location to the rest of system, influencing various types of protections. In addition, since more power-electronics-interfaced generation and energy storage devices are being integrated into power systems, electromechanical wave propagation speeds in the future power systems are likely to change accordingly. In this paper, GPS-synchronized measurement data from a wide-area synchrophasor measurement system FNET/GridEye are used to analyze the characteristics of electromechanical wave propagation in the U.S. Eastern Interconnection (EI) system. Afterwards, high levels of photovoltaic (PV) penetration are modeled in the EI to investigate the influences of a typical power-electronics--interfaced resource on the electromechanical wave propagation speed. The result shows a direct correlation between the local penetration level of inverter-based generation and the electromechanical wave propagation speed.

## I. Introduction

Synchronous generator rotor angles deviate after a disturbance in power systems. The propagation of angle deviation from the disturbance location to the entire synchronized system is referred to as an electromechanical wave. This kind of electromechanical wave has been clearly observed since the deployment of the wide-area synchrophasor measurement technology [1]. Its propagation speed was found to be much smaller than the speed of light, and therefore it was categorized as an intrinsically different wave compared with electromagnetic waves. Electromechanical wave may influence various protections, such as over-current, impedance, and out-of-step relays, as well as under-frequency load-shedding. Meanwhile, the electromechanical wave propagation phenomenon had also been found to be useful in event location [2] and inertia estimation [3].

Currently, there are two categories of methods in general to study electromechanical wave propagation in power systems: simulation-based methods and measurement-based methods. Simulation methods are based on detailed power system dynamic models. Since the system configuration is constantly changing because of unit commitment etc., the simulation method usually uses a typical snapshot of power flow and associated dynamic models as the study case. Comparatively, the measurement-based method is able to calculate electromechanical wave propagation speed using wide-area synchrophasor measurement data during transients.

In the literature, electromechanical wave was mainly studied through simulation-based methods. The continuum model was first proposed in [4] to model uniformly distributed machine inertia and transmission impedance. Based on the continuum model, Ref. [5] derived standard second-order wave equations to describe electromechanical waves. By aggregating local system parameters into cells, Ref. [6] relaxed constraints that the virtual grid is homogenous and isotropic in the continuum model. Ref. [7] further relaxed the assumptions, such as lossless lines and generators. In addition, the Gaussian function was applied to smooth the nodal parameters. Ref. [8] studied the propagation of electromechanical waves based on a reduced model of the U.S. Eastern Interconnection (EI) system. It was found that wave propagation speeds are higher if it propagates along the power transfer direction compared with that of the opposite direction. Some studies developed controllers to mitigate the electromechanical wave based on the wave equation derived from the continuum model [9].

The wide-area synchronized measurements of wave propagation can serve as a foundation for theoretical modeling, validation, and applications development. For example, wave propagation has been observed and studied using data from FNET/GridEye, a wide-area measurement deployed at the distributed level. Ref. [10] proposed a method to calculate regional speeds. This study confirmed that the propagation speed has high correlations with the distribution of system inertia and network impedance. A series of applications including disturbance location have been developed based on electromechanical waves propagation recorded in FNET/GridEye measurements [3, 10, 11].

It was predicted by the ideal continuum model that the changes of power system parameters, such as line impedance and inertia, would influence electromechanical wave propagation speed [12]. As power systems are accommodating more and more power-electronics-interfaced renewable generations and energy storage facilities, system inertia will significantly decrease. However, there has been few literatures that studied the potential change of wave propagation speed due to the increasing renewable generation.

This paper studies the electromechanical wave propagation in current EI system using wide-area measurements and detailed dynamic model simulations were used to provide information on wave propagation in the future system. The rest of paper is organized as follows: Section II introduces the FNET/GridEye measurement system and analyzes the electromechanical wave propagation using FNET/GridEye data in the EI system; Section III studied the impact of high PV penetration on electromechanical wave propagation based on detailed dynamic models of the EI, Section IV concludes this paper.

## II. EI Electromechanical Wave Propagation Analysis Using FNET/GridEye Measurement

### A. FNET/GridEye Overview and Electromechanical Wave Propagation Visualization

The distribution-level synchrophasor technology makes it possible to significantly decrease sensor costs and simplify deployments [13]. Nevertheless, the hardware and software designs for synchrophasor sensors in the distribution system have some special technical challenges [14]. For example, compared with transmission systems, distribution systems have much worse power quality because of harmonics and distortions from various electric appliances. Therefore, the distribution-level sensors should be capable of capturing power grid dynamics at noisy system ends.

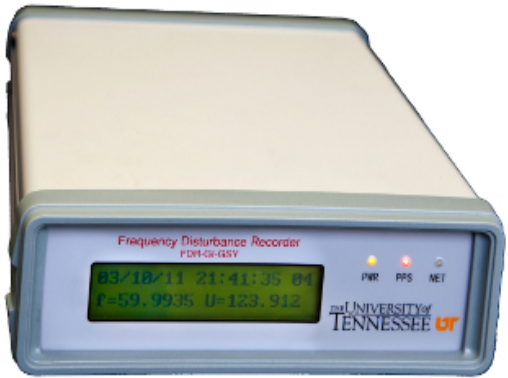

Fig. 1 Generation-II FDR

As the sensor of FNET/GridEye, frequency disturbance recorder (FDR) includes a microprocessor to sample and calculate frequency and voltage phasor, and other modules for GPS time synchronization and Ethernet communication, etc. FDR features a low manufacturing cost, which is about one tenth of a typical phasor measurement unit (PMU) [15, 16]. However, FDR does not sacrifice its accuracy for low cost and quick deployment [17]. In fact, FDR has comparable or even higher accuracy than its counterparts. For example, the expected accuracy of Micro-PMU is ±0.05° for angle measurement, which is much lower than that of FDR [18, 19].

So far, three generations of FDRs have been developed for improved measurement accuracy and data quality. Fig. 1 shows the most widely deployed Generation-II FDR. Updates on Generation-III FDR include enhanced functionalities on power quality measurements and, more importantly, accuracy improvement archived by hardware and algorithm advancements. Its steady-state measurement accuracy reaches a record of ±0.06 mHz for frequency and ±0.003° for angle, respectively.

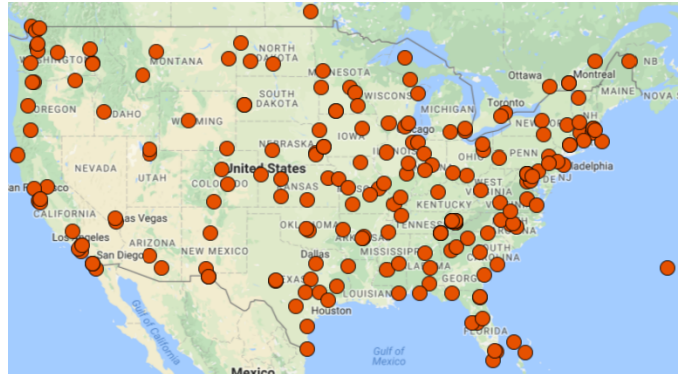

Fig. 2 The FDR location map in North America

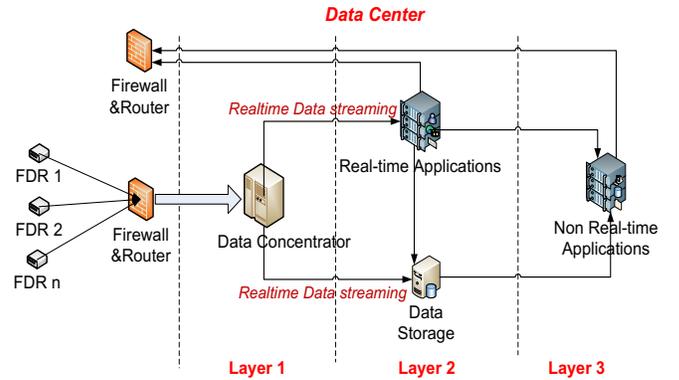

Fig. 3 The FNET/GridEye data centre structure

The deployment of FDR in North America by April 2017 is shown in Fig. 2. Real-time phasor measurements collected by FDRs at the indicated locations are transmitted via Ethernet and collected by the data center hosted at University of Tennessee, Knoxville, and Oak Ridge National Laboratory (ORNL). The data center has a multi-layer structure as shown in Fig. 3 [17]. It is capable of managing, processing, and archiving the large amounts of measurements in a systematical way. The monitoring system enables multiple functionalities including online monitoring, online analysis, and offline data mining. Based on the FNET/GridEye platform, a variety of visualization and analytics applications have been developed, and they are widely adopted by the academia, the industry, and government agencies. These applications enable system operators to be better aware of the spatiotemporal power grid dynamics caused by various disturbances and changing environments.

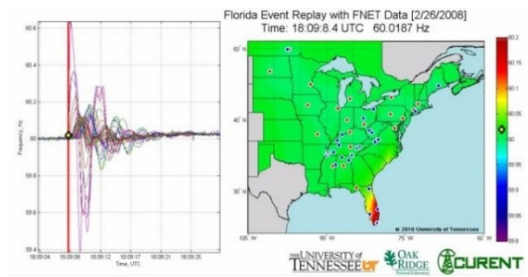

(a) Stage 1

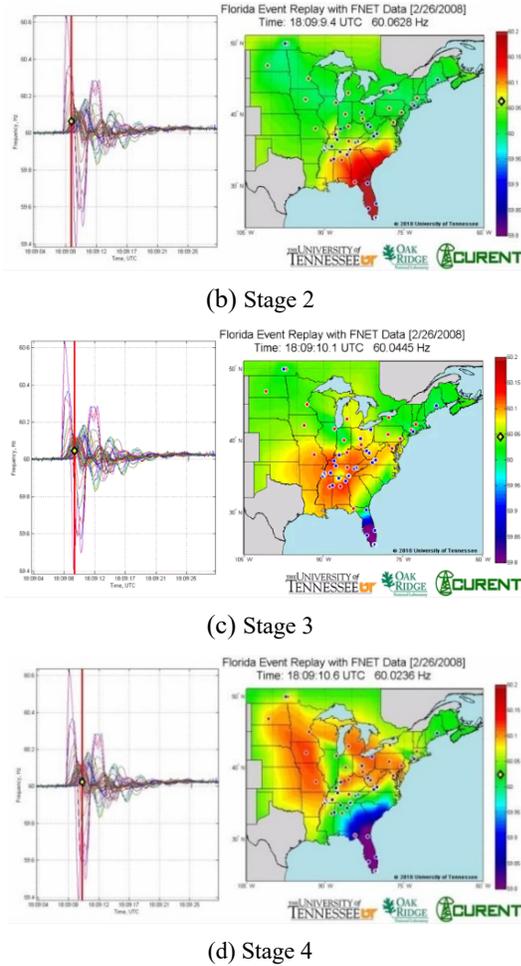

(b) Stage 2

(c) Stage 3

(d) Stage 4

Fig. 4. Electromechanical wave propagation observed in the 2008 Florida blackout event

As one application of FDR measurements, the FNET/GridEye event replay tool can replay the electromechanical wave propagation in any major power system disturbance, one example of which is shown in Fig. 4. This event replay clearly shows the start of the propagation and how the electromechanical wave propagates throughout the EI system.

*B. Electromechanical Wave Propagation Speed Analysis*

The electromechanical wave propagation speed can be quantified by tracing the movement of the wave front geographically. The time delay of the wave front from the disturbance location to an observation location can be calculated by the Time Delay of Arrival (TDOA) of an electromechanical wave. TDOA is defined as the time difference of the disturbance time and the time when the frequency at a given measurement location passing a pre-set threshold.

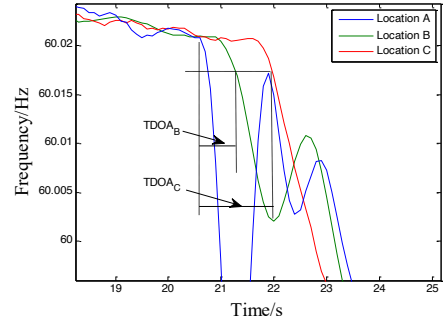

Fig. 5. TDOA calculation

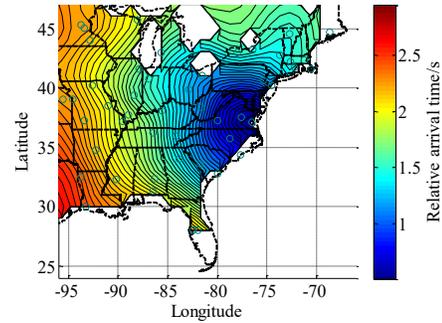

Fig. 6. TDOA distribution map (disturbance in North-eastern U.S.)

Fig. 5 shows the calculation scheme of TDOAs after a generation trip event. Using the TDOA values at different locations across the EI system, a TDOA distribution map can be generated by interpolation as shown in Fig. 6. It shows that the TDOA increases with the distance between the measurement location and the disturbance location.

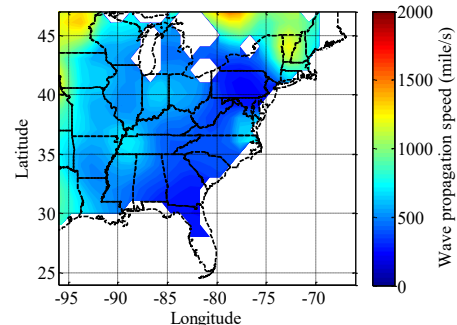

Fig. 7. Electromechanical wave propagation speed distribution (snapshot in a peak load period)

Based on the TDOA distribution map, the wave propagation speed at each location can be calculated by the gradient of the TDOA distribution, as shown in Fig .7. The speed distribution shows that northwestern and northeastern regions have higher wave propagation speeds than the rest of the EI. This is likely due to these regions have fewer generations and loads compared with other parts of the EI.

 

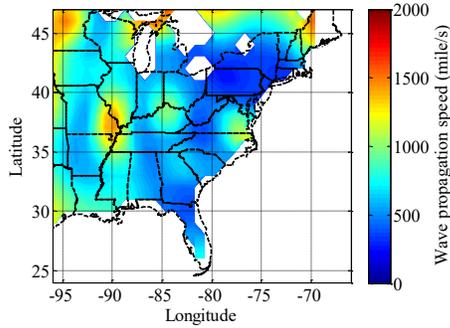

Fig. 8. Electromechanical wave propagation speed distribution (snapshot in an off-peak period)

Fig. 8 shows the wave propagation speed distribution calculated by another generation trip that happened in an autumn off-peak-load period. Compared with the speed distribution in Fig. 7, the wave propagates faster in Fig. 8. The explanation lies in that the summer peak-load period has a larger number of system loads and generations, making it more difficult for electrometrical wave to propagate.

### III. IMPACT OF HIGH PV PENETRATION ON ELECTROMECHANICAL WAVE PROPAGATION

To investigate the impact of high PV penetration on electromechanical wave propagation, the EI MMWG 2030 model [20, 21] is adopted as the study case in this section. The distribution of PV in the high PV EI model was determined by achieving the least cost of system investment and operation.

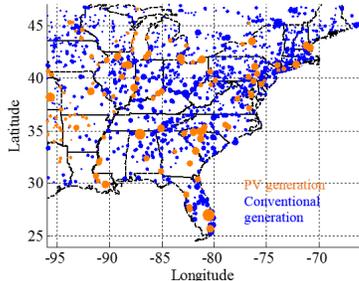

Fig. 9. PV distribution in the 25% PV penetration scenario

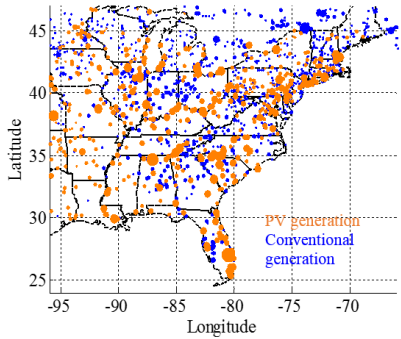

Fig. 10. PV distribution in the 65% PV penetration scenario

The distribution of PV generation and conventional generation for the 25% PV penetration case and the 65% PV penetration case are shown in Fig. 9 and Fig. 10, respectively. It shows that PV penetration levels in southern and northeastern EI regions will be higher than the rest of EI. This indicates that the electrometrical wave propagation speed will increase in those regions.

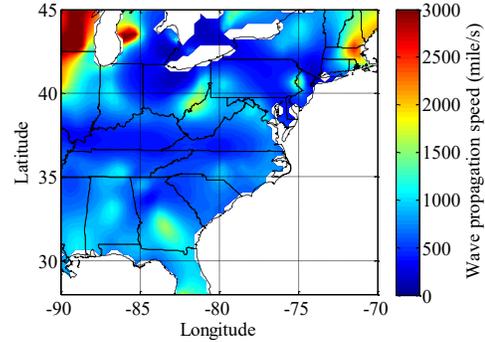

Fig. 11. Electromechanical wave propagation speed distribution in the 25% PV penetration scenario

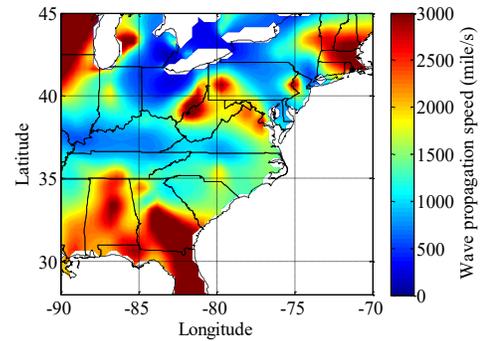

Fig. 12. Electromechanical wave propagation speed distribution in the 65% PV penetration scenario

Using the same approach as described in Section III, the distribution of wave propagation speed for each scenario is presented in Fig. 11 and Fig. 12, respectively. These results demonstrate that the wave propagation speed does increase with the increase of PV penetration. Specifically, comparing Fig.11 with Fig. 7 (note different color scales), 25% PV penetration significantly increases the wave propagation speed due to the decrease of system inertia. Fig. 12 also proves that in some regions with higher PV penetration, such as southern and northeastern EI regions, the electromechanical wave propagates faster at a speed of close or even larger than 3,000 miles per second.

### IV. CONCLUSION

This paper analyzed the electromechanical wave propagation using two approaches: measurement and simulation. It was found by the measurement-based method that the wave propagation has both distinct geographical distribution features and temporal features due to the heterogeneity and temporal changes of the EI system. The simulation-based approach was leveraged to study the change of wave propagation speed due to the integration of more PV generation in the future. The results showed an obvious correlation between PV penetration and wave propagation speed. The electromechanical wave propagation speed may reach 3,000 miles per second for regions with high PV penetration levels.